\documentclass[aps,pre]{revtex4}
\usepackage{amsmath,graphicx}
\usepackage{subfig}
\usepackage{latexsym}
\usepackage{bm}
\usepackage{epstopdf}
\usepackage{amsfonts,paralist,amssymb}
\usepackage{bbold}
\usepackage{stmaryrd}

\def\EE{{\cal E}} 

\newcommand{\beq}{\begin{equation}}
\newcommand{\eeq}{\end{equation}}

\newcommand{\bea}{\begin{eqnarray}}
\newcommand{\eea}{\end{eqnarray}}

\begin{document}
% Title.
% ------
\title{Random Projections through multiple optical scattering:
Approximating kernels at the speed of light}

\author{A. Saade$^1$, F. Caltagirone$^1$, I. Carron $^2$, L. Daudet
  $^{2,3,7}$, A. Dr\'emeau $^4$, S. Gigan $^{5,6,7}$,
  F. Krzakala$^{1,6,7}$}

\affiliation{$^1$ Laboratoire de Physique Statistique, CNRS UMR 8550 \&
  \'Ecole Normale Sup\'erieure, Paris, France.\\
  $^2$Institut Langevin, ESPCI and CNRS UMR 7587, Paris, F-75005,
  France\\
  $^3$Paris Diderot University, Sorbonne Paris Cit\'e, Paris, F-75013,
  France\\
  $^4$ ENSTA Bretagne and Lab-STICC UMR 6285, F-29806 Brest, France \\
  $^5$Laboratoire Kastler Brossel,  CNRS UMR 8552 \&
  \'Ecole Normale Sup\'erieure,75005 Paris, France.\\
  $^6$Sorbonne Universit\'es, Universit\'e Pierre et Marie Curie Paris
06, F-75005,
  Paris, France\\
  $^7$ PSL Research University, 75005 Paris, France}

% \ninept
% 

% 
\begin{abstract}
  Random projections have proven extremely useful in many signal processing and machine
  learning applications.  However, they often require either to store
  a very large random matrix, or to use a different, structured matrix
  to reduce the computational and memory costs.  Here, we overcome this
  difficulty by proposing an analog, optical device, that performs the
  random projections {\it literally} at the speed of light without
  having to store any matrix in memory. This is achieved using the
  physical properties of multiple coherent scattering of coherent light in random media.
We use this device on a simple task of classification with a
  kernel machine, and we show that, on the MNIST database, the experimental results closely match the theoretical performance 
  of the corresponding kernel. 
  This framework can help make kernel methods
  practical for applications that have large training sets and/or
  require real-time prediction. We discuss possible
  extensions of the method in terms of a class of kernels, speed,
  memory consumption and different problems.
\end{abstract}
% 
% \begin{keywords}
%   Kernel methods, random projections,  machine learning, large-scale data processing, 
%   optical computing
% \end{keywords}

\maketitle

\section{Introduction}
\label{sec:intro}

Random projections have proven useful in signal processing and machine learning in several
ways.
% \cite{candes2006stable, donoho2006compressed, meng2013low,
% paul2012random, durrant2014random, fern2003random}
 A first line of applications is concerned with dimensionality
reduction
\cite{bingham2001random,fradkin2003experiments,achlioptas2003database,hegde2008random}. Given
a dataset with a very large number of features, we look for a simple
transformation of the data that reduces the number of features while
approximately preserving the pairwise distances between data
points. It turns out that linear random projection are suitable to
this purpose (Johnson-Lindenstrauss lemma
\cite{johnson1984extensions}). Similarly, a small number of random projections of sparse signals can
carry sufficient information for their reconstruction, as shown in compressed sensing \cite{candes2006stable, donoho2006compressed}.

% Similarly, a small number of random projections of 
% sparse signals can carry sufficient information for their reconstruction, as 
% shown  by the compressed sensing theory.   

Conversely, non-linear random projections allow the embedding of a dataset in
a larger dimensional feature space. Notably, the data may become
linearly separable in this larger space, while it was not in the native
feature space. A linear model can then be used to fit or classify the
data. An interesting class of methods relying on this embedding is
that of so-called kernel machines \cite{shawe2004kernel}, among which
is the celebrated SVM. These offer the advantage, sometimes called
\emph{kernel trick}, of not requiring the explicit mapping, but only
the inner products of all pairs of embedded data points, i.e. a
function $k(x,y)$ where $k$ is called the kernel, and $x,y$ are data
points in the native feature space. While removing the dependency on
the dimension of the embedding (which is possibly infinite), this
trick relies on a matrix storing all the values of $k$ for all pairs
of data points. This typically does not scale to large datasets.

Recently \cite{rahimi2007random}, a cure has been proposed to this
problem that actually makes explicit a non-linear mapping of the data
points such that the inner product of two transformed data points
approximately equals their kernel evaluation. In this new feature
space, the regression or classification task can be solved by a linear
machine, which can be trained very quickly compared to non-linear ones
\cite{joachims2006training}. Random projections have thus made some large-scale machine learning 
problems tractable, to the
point that actually computing them - i.e. generating and storing a
large random matrix, and computing matrix-vector products - has become
one of the major bottlenecks in the above mentioned approaches \cite{ailon2009fast, le2013fastfood, meng2013low}.

In this study, we demonstrate experimentally that an optical-based
hardware can be built which instantaneously provides a large number of
``ideal'' random projections of any input data. We show that this
physical process mimics, for the goal of classification, the
computation of a well-defined elliptic kernel. In other words, we
believe that this optical setup can be seen as a generic digital data
pre-processor, for many subsequent uses of the data. It has to be
emphasized that, contrary to previous studies on \textit{imaging}
techniques based on multiple scattering (with a similar experimental
setup \cite{liutkus2014imaging,dremeau2015reference}), this random
transformation does not have to be precisely determined through a lengthy
calibration stage : the knowledge of its statistical properties
is enough to guarantee its effectiveness.
% , as long as the medium remains stable over time. 
% Importantly, this random transformation does not change over time (the medium is stable). 
Although this process is fundamentally
analog, we show here that the associated experimental noise does not
substantially change the results, which are in very good agreement with computer simulations.  
% as compared to the same computations performed on a standard computer.

\section{Random projections and kernel machine learning}
Let us start by introducing the standard ridge regression problem,
arguably the simplest kernel method for supervised classification.
Throughout the section, we are given a training set composed of
labeled data, represented by a matrix
${\bf U} \in \mathbb{R}^{n\times p}$ where $n$ is the number of
samples and $p$ is the dimension of the data (number of features).
% , and by a matrix ${\bf Y} \in \mathbb{R}^{n \times q}$ encoding the labels. 
Each data point ${\bf U}_i\in\mathbb{R}^p$ has a unique label $l_i\in\llbracket 1,q\rrbracket$ which we encode in a matrix ${\bf Y} \in \mathbb{R}^{n \times q}$ with elements ${\bf Y}_{i,j} = \delta_{j,l_i}$.
We are also given a test set of unlabeled data, represented by a matrix
${\bf \tilde{U}} \in \mathbb{R}^{\tilde{n}\times p}$ where $\tilde{n}$ is the
number of unlabeled samples. Our goal is to estimate their label $\tilde{{\bf Y}}\in \mathbb{R}^{\tilde{n} \times q}$ using a classifier trained on the training set. Recall that the ridge regression problem reads 
\beq
\label{ridge regression}
\underset{\beta \in \mathbb{R}^{p \times q} }{\rm argmin}\
||{\bf U}\beta - {\bf Y}||_2^2 + \gamma ||\beta||_2^2 
\eeq
where $\gamma$ controls the trade-off between the estimation error and the regularization. Its closed-form solution is given by 
\beq
\beta=({\bf U}^T{\bf U}+\gamma {\bf I}_p)^{-1} {\bf U}^T {\bf Y} = {\bf U}^T ({\bf U}{\bf U}^T+\gamma {\bf I}_n)^{-1}  {\bf Y}
\eeq
where we have used a classical algebraic identity.
Our prediction for the labels of the test data points is then
\begin{align}
\label{notgram}
{\bf \tilde{Y}} = {\bf \tilde{U}}\beta &={\bf \tilde{U}} ({\bf U}^T{\bf U}+\gamma {\bf I}_p)^{-1} {\bf U}^T{\bf Y}\\
&= {\bf \tilde{U}}{\bf U}^T ({\bf U}{\bf U}^T+\gamma {\bf I}_n)^{-1}  {\bf Y} 
\label{gram}
\end{align}
More precisely, since we are interested in classification, we will set the label of ${\bf \tilde{U}}_i$ to $\tilde{l}_i = \max_{j\in\llbracket 1,q \rrbracket} {\bf \tilde{Y}}_{ij}$. 
Note that Eq. (\ref{gram}) only depends on inner products of the data points.

\subsection{Kernel classification}

Given a kernel $k$, we define the $n \times n$ (resp. $\tilde{n}\times n$) kernel matrix ${\bf K}$ (resp. ${\bf \tilde{K}})$ with elements 
\begin{align}
{\bf K}_{i,j} =k({\bf U}_{i},{\bf U}_{j}) ~~\text{and}~~{\bf \tilde{K}}_{i,j} =k({\bf \tilde{U}}_{i},{\bf U}_{j})
\end{align}
In the kernel ridge regression problem, we simply replace the Euclidean inner product in feature space by this kernel. Because Eq. 
(\ref{gram}) only depends on these inner products, we do not have to
explicit the mapping, and the solution of the kernel ridge regression
is directly given by \beq \tilde{\bf Y} =\tilde{\bf K} ( {\bf K} +
\gamma {\bf I}_n)^{-1} {\bf Y}
\label{def:kernel_classif}
\eeq
We consider here the following elliptic kernel: 
% for ${\bf U}_i,{\bf U}_j\in\mathbb{R}^p$
%
%\bea
\begin{align}
 k({\bf U}_i,{\bf U}_j)&=\frac{\sqrt{{\bf U}_i^T{\bf U}_i \,
    {\bf U}_j^T{\bf U}_j}}{2} \Bigg\{ -
(\sin^2\theta)\EE_K\left[
  \cos^2\theta \right] +2 \EE_E\left[ \cos^2\theta \right] \negmedspace+\negmedspace|\sin{\theta}|\left(2 \EE_E\negmedspace\left[
  -\frac{\cos^2\theta}{\sin^2\theta} \right]\negmedspace -\negmedspace \EE_K\negmedspace\left[
  -\frac{\cos^2\theta}{\sin^2\theta}\right ] \right)\!\!\!\Bigg\}
\label{kernel_Elliptic}
\end{align}
%\eea
%
where $\EE_K[.]$ and $\EE_E[.]$ are the complete elliptic integrals of the first and second kind respectively,
and $\theta$ the angle between ${\bf U}_i$ and ${\bf U}_j$. Despite its apparent
complexity, this function essentially looks like a bell curve as a
function of $\theta$, and therefore quantifies the similarity between ${\bf U}_i$ and ${\bf U}_j$. This particular choice will be motivated experimentally in section \ref{complex-proj-elliptic}.

To illustrate this method, we consider the MNIST dataset of
handwritten digits \cite{lecun1998gradient}.  This dataset is split
into a training set of $60 000$ digits of size $28\times28$, and a
test set of $10 000$ digits.  Using the elliptic kernel, we achieve a
classification error of $1.31\%$, to be compared with a $12\%$ error
for the purely linear ridge regression (\ref{gram}). While better
results can be achieved using much more complex deep neural networks,
kernel ridge regression achieves a reasonable error and is remarkable
in its simplicity. The major drawback of this approach is the
computational cost of inverting an $n\times n$ matrix. In ``big data''
problems, where $n$ can be in the billions, it would not even be
possible to store such a matrix.

\subsection{Random projections}

Following \cite{rahimi2007random}, we now compute a non-linear mapping
and solve the linear ridge regression problem (\ref{ridge regression}) in this new feature
space of dimension $N$. Specifically, we consider mappings of the form
\beq
{\bf X}_{i,j} = \phi ( ({\bf W U}_{i})_j + {\bf b}_j)\qquad i\in\llbracket1,n\rrbracket,j\in\llbracket1,N\rrbracket
% \qquad 1\leq i\leq N,1\leq k\leq p
\label{projection}
\eeq
where ${\bf b}$ is a bias, $\phi$ is a non-linear function and ${\bf W}\in\mathbb{R}^{N\times p}$ is a
linear projection which we take to be random. The solution of the
ridge regression problem is  
\beq 
\tilde{\bf Y} \negmedspace=\negmedspace {\bf
  \tilde{X}}{\bf X}^T\negmedspace ({\bf X}{\bf X}^T\negmedspace+\negmedspace\gamma {\bf I}_n)^{-1} {\bf Y} \negmedspace= \negmedspace{\bf \tilde{X}} ({\bf X}^T{\bf X}\negmedspace+\negmedspace\gamma {\bf I}_N)^{-1} {\bf X}^T {\bf Y}\negmedspace
\label{projected ridge}
\eeq 
In this last formulation, we see that if the inner products of
the data points in the new feature space approximate a kernel, and if
this new feature space has a reasonable dimension $N$, then we can
approximate a kernel ridge regression by inverting a smaller matrix (${\bf X}^T{\bf X}\in\mathbb{R}^{N\times N}$) instead of the kernel
matrix (${\bf X}{\bf X}^T\in\mathbb{R}^{n\times n}$). Concretely, the experimental device described in the next section can perform
(\ref{projection}) with ${\bf W}$ a random i.i.d complex matrix with
Gaussian real and imaginary parts, $\phi$ the modulus
function, and a vanishing bias ${\bf b}$. This setting has been popularized recently under the name Extreme Learning Machine \cite{huang2006extreme,huang2004extreme}. In synthetic experiments, if we use $N=10 000$ random
projections and the linear ridge regression (\ref{projected ridge}), we
get an error of approximately $2\%$ (see Fig.~\ref{fig:results}),
having only to invert a $10 000^2$ matrix instead of a $60 000^2$
one. Even better: now we have lost the dependency on $n^2$.
\subsection{Random complex projections and the elliptic kernel}

%
% The link between the two methods can be understood right away using
% the classical identity
% % 
% \beq (XX'+\gamma {\bf I}_n)^{-1} X = X'(X'X+\gamma {\bf I}_N)^{-1}
% \eeq
% %
% with which we can rewrite
% %
% \beq \beta=X'(X'X+\gamma {\bf I}_N) {\bf Y} \eeq
% %
% Therefore we have, as in eq. (\ref{def:kernel_classif}):
% % 
% \beq
% \tilde{\bf Y} =\tilde{\bf K} ( {\bf K} + \gamma {\bf I_N})^{-1} {\bf
%   Y}\eeq
% %
% where this time, however, the matrices {\bf K} are defined by
% \beq
% \tilde {\bf K}=\tilde{\bf X'}{\bf X}~~~ \text{with}~~~{\bf K}={\bf X'}{\bf X}
% \eeq
% %
% We thus replace the kernel Gram matrix by using the scalar product
% between the random non linear projections. When the number of
% projections grows, thanks to the concentration of the measure, the
% scalar product tends to a kernel function that depends only on the
% values of ${\bf U}$ \cite{williams1998computation}.
\label{complex-proj-elliptic}

\begin{figure}[t!]
  \centering \centerline{\includegraphics[scale=1]{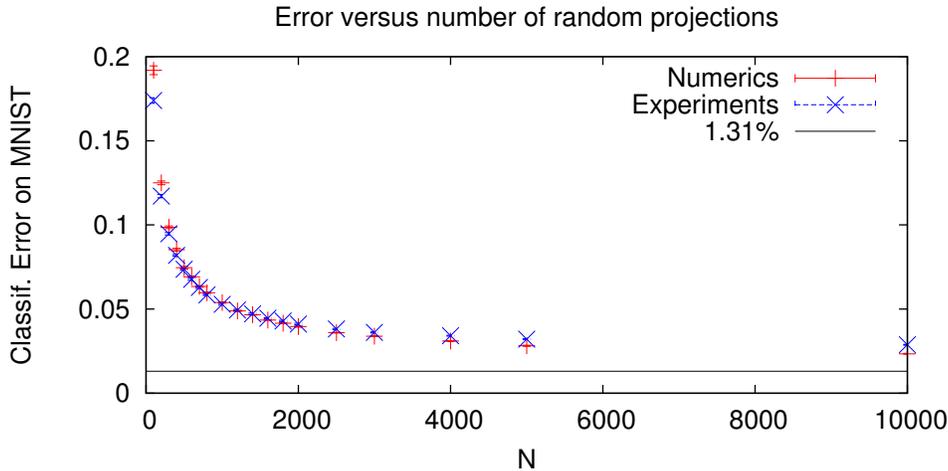}}
  \caption{Comparison between synthetic random projections and the
    experimental ones. The classification error is shown as a function
    of the dimension $N$ of the feature space. As $N$ increases, the
    error using synthetic random projections (red) approaches the
    asymptotic value using the true elliptic kernel (here $1.31
    \%$).
    Experimental results, made with our optical device, closely follow
    the synthetic ones though they tend to deviate from the synthetic
    data for larger $N$ (see discussion).}
\label{fig:results}
  \end{figure}

In the limit where the number of projections $N$ grows to infinity, due to the concentration of the measure,
it is possible to show that the inner product between the projected data points ${\bf X}_i\in\mathbb{R}^N$ tends to a kernel function that depends only on the ${\bf U}_i$ \cite{williams1998computation}. With the choice of ${\bf W}$ and $\phi$ corresponding to our experiment, a tedious but simple computation shows that the limiting kernel is precisely the elliptic kernel of equation (\ref{kernel_Elliptic}). To approximate this kernel, we can therefore use our experimental device to perform the non-linear projections, and solve the ridge regression problem (\ref{projected ridge}). This concentration phenomenon is shown on  Fig.~\ref{fig:results} where we plot the classification error on the MNIST database when the number of random projections
is increased (red points). As the number of random projections grows, the classification error using synthetic random projections
approaches the asymptotic value using the true elliptic kernel
($1.31 \%$).  Empirically, we find that these corrections are very well
fitted by a power law in $N^{-2/3}$.

% In particular, it is possible to show with a tedius but simple
% computation \cite{KERNEL} that, when using complex random projections
% and the modulus function for $\phi$, the values of
% ${\bf X_1.}{\bf X_2}$ converges towards a kernel function $K(U_1,U_2)$
% which is nothing but the elliptic kernel! This is why we considered
% this choice in the present contribution. This concentration phenomena
% is illustrated in Fig.~\ref{fig:results} where we plot our prediction
% abilities on the MNIST database when the number of random projections
% is increased (red points).  As the number of random projections grows
% the error classification using synthetic random projections (red)
% approaches the asymptotic value using the true Elliptic kernel
% ($1.31 \%$).  Empirically, we find these corrections follow very well
% a power low in $N^{-2/3}$.

% -------------------------------------------------------------------------

\section{Experimental Apparatus}

The approach exposed in the previous section still requires to store, and multiply by, a potentially huge 
random matrix, and to apply the modulus function. We now move to the
discussion of our experimental apparatus which will make this
procedure a trivial, instantaneous one.

% -------------------------------------------------------------------------
\begin{figure}[ht]
\centering
\centerline{\includegraphics[scale=0.6]{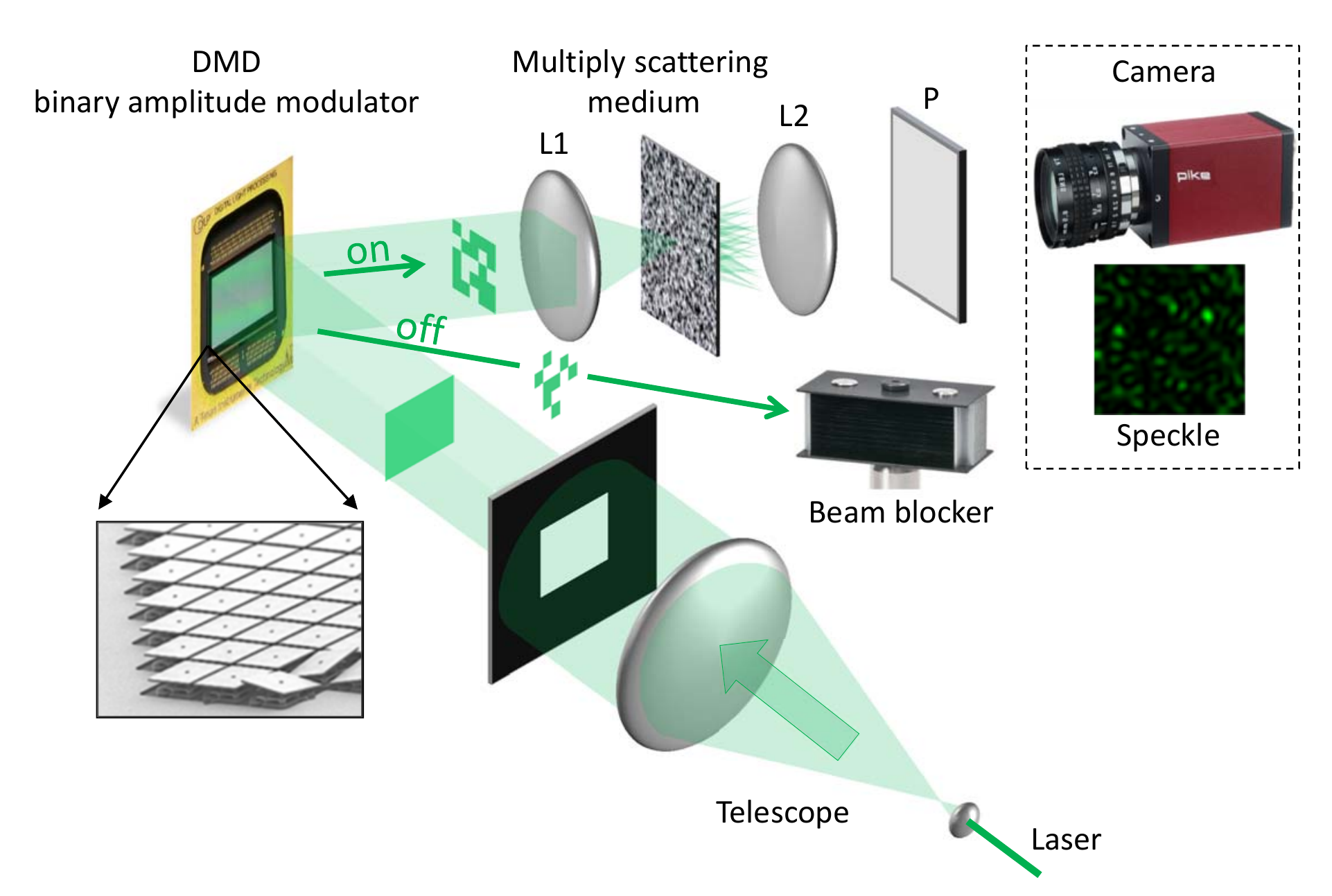}}
\caption{Experimental setup (from  \cite{dremeau2015reference}). A monochromatic laser at 532nm is
  expanded by a telescope, then illuminates a digital micromirror device 
  (DMD), able to spatially encode digital information on the light
  beam by amplitude modulation,
  %The DMD
  %and the encoding of the information are 
  as described in section \ref{encoding}. 
  The light beam carrying the signal is then focused on a
  random medium by means of a lens. Here, the medium is a thick
  (several tens of microns) layer of $TiO_2$ (Titanium Dioxide)
  nanoparticles (white paint pigments deposited on a microscope glass
  slide). The transmitted light is collected on the far side by a
  second lens, passes through a polarizer, and is measured by a
  standard monochrome CCD camera. }
\label{fig:experiments}
  \end{figure}
\subsection{Principle of optical analog random projections}

At the basis of the analog random projections, we exploit the ability
of a heterogeneous material, such as paper, paint, or any white
translucid material, to scatter light impinging on them in a very
complex way. Due to their extremely high complexity, their behavior
for light scattering is considered ``random'', and this kind of medium
are often called ``random medium''.  This term is abusive since
scattering is a linear, deterministic, and reproducible
phenomenon. However, the unpredictable nature of the process makes it
effectively a random process. This is why these materials are called
``opaque'', since all information on the incoming light is seemingly
lost (but only irremediably mixed) during the propagation.
As an example, consider a cube of edge size 100 $\mu$m. It comprises
$\approx 10^7$ paint nanoparticles, whose positions and shape would have to
be exactly known in order to predict its effect on light. Propagation through such a layer 
can be seen as a
random walk because of frequent scattering  with the
nanoparticles. The characteristic step is the transport mean free
path, typically of a few $\mu$m, so light would explore the whole
volume and endure on average tens of thousands of such steps before
exiting on the other side, in a few picoseconds.

When light is coherent, it gives rise to interferences. The complex
interference pattern arising from the multiple scattering process is
called ``speckle''. It is characteristic not only of the heterogeneous
material, but also of the ``shape'' of the input light. In essence, the
propagation of light through a random medium can be modeled as $y={\bf H}x$
where $y$ (resp. $x$) are vector amplitudes between a set of spatial
modes at the output (resp. input) of the medium, and where ${\bf H}$ is the
so-called transmission matrix of the medium, which has been shown to
be very close to Gaussian i.i.d matrices \cite{popoff2010measuring}. By
``interrogating'' the medium with an appropriate set of input
illuminations, which can be conveniently done using a spatial light
modulator and a laser, and measuring the resulting ``answer'' by means
of a CCD or CMOS camera for instance, one would therefore record the
resulting intensity $\lvert y \rvert^2$ (see Fig.~\ref{fig:experiments}). For a stable medium, such as
a paint layer for instance, the transmission matrix ${\bf H}$ is stable and
the medium can therefore provide a convenient platform for random
projections, without the need to determine ${\bf H}$. 

\subsection{Encoding the data on the DMD}
\label{encoding}
The DMD %(see Fig. \ref{fig:experiments}) 
we use is a $1920\times1080$ array of 
micro-mirrors. Each micro-mirror encodes a binary value (lit or
not). Levels of grey can be conventionally coded in time by lighting up a micro-mirror for a
fraction of the exposure time of the camera. This approach however reduces the
frequency at which data can be processed through the apparatus. We
therefore resorted to spatial encoding. We encoded each pixel of
an image on a $4\times4$ array of micro-mirrors in which the number of
lit micro-mirrors reflects the level of grey of the pixel. This
allowed us to encode $17$ grey levels.  

Each digit in the MNIST dataset can be seen as a $28^2$ array of
integers between $0$ and $255$. We first quantized 
the grey levels %it to turn it into a $28^2$ array of integers 
between $0$ and $16$. 
We then encoded each
of these quantized pixels as a $4\times4$ array of micro-mirrors. 
%, as described above. 
This results in a $112^2$ array of binary variables.
To rescale the resulting image to $1920\times1080$, which is the size
of the DMD, we need to first add a border of zeros, so that the image
becomes of size $120^2$. Finally, we can rescale the image by a factor
of $16\times9$. The result is a $1920\times1080$ binary image,
encoding each digit over $17$ levels of grey. This image can then be
projected on the DMD. It is in principle possible to encode more
levels of grey by encoding each pixel on a larger array of
micro-mirrors. 
%The resulting image is however bigger, which slows down
%the processing of the data, compromising the stability of the
%disordered medium over the duration of the experiment. These problems
%are readily solved by a faster apparatus, or a more stable medium.
The procedure described to encode the data on the DMD is not specific
to images and generalizes to any kind of data which can be written as
a numerical vector.

After sending the data through the disordered medium, we acquire a
snapshot of the resulting random projection using a standard
camera. To reduce the correlations between neighboring pixels of the
output, we start by acquiring a $400^2$ pixel area, which we then reduce
to $100^2$ pixels where each of these pixels is the average of a $4^2$
patch in the original snapshot.

\section{Results}
Our results are summarized in Fig.~\ref{fig:results} where we compare
the efficiency in classification using synthetic random projection (in
red) and the random projection obtained within the experimental set-up
(in blue). We find that the two curves agree remarkably well, thus
validating our procedure to generate analog random projections for
classification tasks.

We note, however, that when the number of projections is increased to
large values, i.e. $N \gtrsim 2000$, the optical experiment starts to
deviate from the synthetic one.  We checked numerically that adding
even large noise to the synthetic projection did not change the
results of the classification, which is therefore robust to
experimental noise.  We thus believe that this deviation is due to residual correlations between pixels of the output image. 
% The rank of the
% random transmission matrix which models the action of our device is a
% function of the number of \emph{free optical modes}
% \cite{popoff2010measuring}. 
The number of independent pixels can be tuned in several ways,
for instance by ensuring that the CCD pixel on the output camera
matches the size of a speckle grain, or by tuning the thickness of the
random medium itself to increase the number of independent modes \cite{popoff2010measuring}. We are therefore confident that this deviation
can be decreased with further development of the
experimental apparatus.

\section{Conclusion and perspectives}
This study shows that large dimensional (here $10^4$ - $10^6$) random projections 
of digital data can be obtained almost
instantaneously with an optical device, that can then be used for
practical machine learning applications such as (but not limited to)
kernel classification. The speed of ``computation'' is only limited by
how fast one can modulate light at the input (off-the-shelf DMDs can
handle images with millions of pixels, at rates of $20$ kHz or above),
and measure the corresponding scattered light.
% (there is a tradeoff
%here in the number of output features that are desired : single pixel
%detectors can sample at much faster rates than the DMD, larger
%detector arrays get slower). 
Note also that although optical sensors natively measure only the
intensity (squared modulus) of the field, and not the complex-valued
field itself, this is precisely what is needed for the
kernel computations presented here, akin to the non-linear transform
of each layer of a neural network. In the cases where the 
linear projections are needed, e.g. for randomized linear algebra, the
experimental setup can be modified to include an interferometric
arm. We can then apply any non-linear function using e.g. FPGAs, widening the range of kernels we can approximate.
It is also possible to combine several DMDs together to process  larger signals.

More generally, we believe that this experiment is a first proof-of-concept toward a new generation of analog optical-based
co-processors, able to complement existing silicon-based chips for the
processing of very large datasets, with potential benefits in terms of
data throughput and energy consumption.

\section*{Acknowledgments}
We thank Gilles Wainrib for illuminating discussions. This research
%leading to these results 
has received funding from the European
Research Council under the EU's $7^{th}$ Framework
Programme (FP/2007-2013/ERC Grant Agreement 307087-SPARCS and
278025-COMEDIA) ; and from LABEX WIFI under 
references ANR-10-LABX-24 and ANR-10-IDEX-0001-02-PSL*.

% \vfill\pagebreak

\bibliographystyle{plain}
\bibliography{mybib}

\begin{thebibliography}{10}

\bibitem{achlioptas2003database}
Dimitris Achlioptas.
\newblock Database-friendly random projections: Johnson-lindenstrauss with
  binary coins.
\newblock {\em Journal of computer and System Sciences}, 66(4):671--687, 2003.

\bibitem{ailon2009fast}
Nir Ailon and Edo Liberty.
\newblock Fast dimension reduction using rademacher series on dual bch codes.
\newblock {\em Discrete \& Computational Geometry}, 42(4):615--630, 2009.

\bibitem{bingham2001random}
Ella Bingham and Heikki Mannila.
\newblock Random projection in dimensionality reduction: applications to image
  and text data.
\newblock In {\em Proceedings of the seventh ACM SIGKDD international
  conference on Knowledge discovery and data mining}, pages 245--250. ACM,
  2001.

\bibitem{candes2006stable}
Emmanuel~J Candes, Justin~K Romberg, and Terence Tao.
\newblock Stable signal recovery from incomplete and inaccurate measurements.
\newblock {\em Communications on pure and applied mathematics},
  59(8):1207--1223, 2006.

\bibitem{donoho2006compressed}
David~L Donoho.
\newblock Compressed sensing.
\newblock {\em Information Theory, IEEE Transactions on}, 52(4):1289--1306,
  2006.

\bibitem{dremeau2015reference}
Ang{\'e}lique Dr{\'e}meau, Antoine Liutkus, David Martina, Ori Katz, Christophe
  Sch{\"u}lke, Florent Krzakala, Sylvain Gigan, and Laurent Daudet.
\newblock Reference-less measurement of the transmission matrix of a highly
  scattering material using a dmd and phase retrieval techniques.
\newblock {\em Optics express}, 23(9):11898--11911, 2015.

\bibitem{fradkin2003experiments}
Dmitriy Fradkin and David Madigan.
\newblock Experiments with random projections for machine learning.
\newblock In {\em Proceedings of the ninth ACM SIGKDD international conference
  on Knowledge discovery and data mining}, pages 517--522. ACM, 2003.

\bibitem{hegde2008random}
Chinmay Hegde, Michael Wakin, and Richard Baraniuk.
\newblock Random projections for manifold learning.
\newblock In {\em Advances in neural information processing systems}, pages
  641--648, 2008.

\bibitem{huang2004extreme}
Guang-Bin Huang, Qin-Yu Zhu, and Chee-Kheong Siew.
\newblock Extreme learning machine: a new learning scheme of feedforward neural
  networks.
\newblock In {\em Neural Networks, 2004. Proceedings. 2004 IEEE International
  Joint Conference on}, volume~2, pages 985--990. IEEE, 2004.

\bibitem{huang2006extreme}
Guang-Bin Huang, Qin-Yu Zhu, and Chee-Kheong Siew.
\newblock Extreme learning machine: theory and applications.
\newblock {\em Neurocomputing}, 70(1):489--501, 2006.

\bibitem{joachims2006training}
Thorsten Joachims.
\newblock Training linear svms in linear time.
\newblock In {\em Proceedings of the 12th ACM SIGKDD international conference
  on Knowledge discovery and data mining}, pages 217--226. ACM, 2006.

\bibitem{johnson1984extensions}
William~B Johnson and Joram Lindenstrauss.
\newblock Extensions of lipschitz mappings into a hilbert space.
\newblock {\em Contemporary mathematics}, 26(189-206):1, 1984.

\bibitem{le2013fastfood}
Quoc Le, Tam{\'a}s Sarl{\'o}s, and Alex Smola.
\newblock Fastfood-approximating kernel expansions in loglinear time.
\newblock In {\em Proceedings of the international conference on machine
  learning}, 2013.

\bibitem{lecun1998gradient}
Yann LeCun, L{\'e}on Bottou, Yoshua Bengio, and Patrick Haffner.
\newblock Gradient-based learning applied to document recognition.
\newblock {\em Proceedings of the IEEE}, 86(11):2278--2324, 1998.

\bibitem{liutkus2014imaging}
Antoine Liutkus, David Martina, S{\'e}bastien Popoff, Gilles Chardon, Ori Katz,
  Geoffroy Lerosey, Sylvain Gigan, Laurent Daudet, and Igor Carron.
\newblock Imaging with nature: Compressive imaging using a multiply scattering
  medium.
\newblock {\em Scientific reports}, 4, 2014.

\bibitem{meng2013low}
Xiangrui Meng and Michael~W Mahoney.
\newblock Low-distortion subspace embeddings in input-sparsity time and
  applications to robust linear regression.
\newblock In {\em Proceedings of the forty-fifth annual ACM symposium on Theory
  of computing}, pages 91--100. ACM, 2013.

\bibitem{popoff2010measuring}
SM~Popoff, G~Lerosey, R~Carminati, M~Fink, AC~Boccara, and S~Gigan.
\newblock Measuring the transmission matrix in optics: an approach to the study
  and control of light propagation in disordered media.
\newblock {\em Physical review letters}, 104(10):100601, 2010.

\bibitem{rahimi2007random}
Ali Rahimi and Benjamin Recht.
\newblock Random features for large-scale kernel machines.
\newblock In {\em Advances in neural information processing systems}, pages
  1177--1184, 2007.

\bibitem{shawe2004kernel}
John Shawe-Taylor and Nello Cristianini.
\newblock {\em Kernel methods for pattern analysis}.
\newblock Cambridge university press, 2004.

\bibitem{williams1998computation}
Christopher~KI Williams.
\newblock Computation with infinite neural networks.
\newblock {\em Neural Computation}, 10(5):1203--1216, 1998.

\end{thebibliography}

\end{document}